\documentstyle[11pt,menu99,epsfig]{article}

\begin{document}
    \setlength{\baselineskip}{2.6ex}


\font\fifteen=cmbx10 at 15pt
\font\twelve=cmbx10 at 12pt

\begin{titlepage}

\begin{center}

\renewcommand{\thefootnote}{\fnsymbol{footnote}}

{\twelve Centre de Physique Th\'eorique\footnote{
Unit\'e Propre de Recherche 7061
}, CNRS Luminy, Case 907}

{\twelve F-13288 Marseille -- Cedex 9}

\vspace{1 cm}

{\fifteen Electromagnetic Corrections to Low-Energy {\Large{\mbox{$\pi-\pi$}}}
Scattering\footnote{Work supported in part by TMR, EC-contract No. ERBFMRX-CT980169 (EURODAPHNE)} 
\\
}

\vspace{0.3 cm}

\setcounter{footnote}{0}
\renewcommand{\thefootnote}{\arabic{footnote}}

{\bf 
Marc KNECHT\footnote{
E-mail address:~knecht@cpt.univ-mrs.fr}
}

\vspace{2,3 cm}

{\bf Abstract}

\end{center}

Electromagnetic corrections to the low-energy  
$\pi^+\pi^-\to\pi^0\pi^0$ scattering amplitude  at 
next-to-leading order in the chiral expansion are reviewed. Their effects on 
the corresponding scattering lengths are estimated and compared 
to the two-loop strong interaction contributions.

\vspace{6 cm}


\bigskip


\bigskip

\bigskip

\bigskip

\bigskip

\noindent
{\it Contribution to the} Eighth International Symposium on Meson-Nucleon Physics and the Structure of the Nucleon, {\it Zuoz, Engadine, Switzerland, August 15-21 1999}

\bigskip

\bigskip

\bigskip

\bigskip

\bigskip

\bigskip

\bigskip

\bigskip

\noindent December 1999

\noindent CPT-99/P.3905

\bigskip

\noindent anonymous ftp or gopher: cpt.univ-mrs.fr

\renewcommand{\thefootnote}{\fnsymbol{footnote}}

\end{titlepage}

\begin{titlepage}

$\ $

\end{titlepage}

\setcounter{footnote}{0}
\renewcommand{\thefootnote}{\arabic{footnote}}


\title{Electromagnetic Corrections to Low-Energy $\pi\pi$ Scattering}
\author{M. Knecht\thanks{Work supported in part by TMR, EC-contract No. ERBFMRX-CT980169 (EURODAPHNE).} \\
{\em Centre de Physique Th\'eorique, CNRS Luminy, Case 907\\ 
F-13288 Marseille Cedex 9, France}} 

\maketitle

\begin{abstract}
\setlength{\baselineskip}{2.6ex}
Electromagnetic corrections to the low-energy  
$\pi^+\pi^-\to\pi^0\pi^0$ scattering amplitude  at 
next-to-leading order in the chiral expansion are reviewed. Their effects on 
the corresponding scattering lengths are estimated and compared 
to the two-loop strong interaction contributions. 

\end{abstract}

\setlength{\baselineskip}{2.6ex}

\section*{INTRODUCTION}

The Chiral Perturbation Theory (ChPT) analyses, in both the 
generalized~ \cite{pipigen1,pipigen2} and the 
standard~ \cite{pipistd1,pipistd2} 
frameworks, of two-loop effects in low-energy 
$\pi\pi$ scattering lead to 
strong interaction corrections which are rather small as compared to the 
leading order and one-loop contributions, of the order of 5\% in the case of 
the S wave scattering lengths $a_0^0$ and $a_0^2$, for instance. 
This leads one to expect that 
higher order corrections to these quantities are well 
under control and can be safely neglected. However, these calculations were 
undertaken without taking into 
account isospin breaking effects, coming either from the mass difference 
between the $u$ and $d$ quarks, or from the electromagnetic interaction. The 
smallness of the two-loop corrections naturally raises the question of how 
they compare to these isospin breaking effects. The quark mass difference 
induces corrections of the order ${\cal O}\big((m_d-m_u)^2\big)$, which are 
expected to be negligible, as already known to be the case for the pion mass 
difference $M_{\pi^{\pm}} - M_{\pi^0}$, the latter being in fact 
dominated by electromagnetic effects due to the virtual photon cloud~
 \cite{GLmasses}. 

In the present contribution, we shall review the status of radiative 
corrections to the amplitude $\pi^+\pi^-\to\pi^0\pi^0$~ \cite{KneUrech}. 
The reason why we focus on the latter comes from the fact that it directly 
appears in the expression of the lifetime of the $\pi^+\pi^-$ dimeson 
atom~ \cite{JalSazd,Ivanov,Soto,Gall} (see in particular the last of these 
references), that will be measured by the DIRAC experiment at 
CERN~ \cite{dirac,diraczuoz}.

\section*{VIRTUAL PHOTONS IN ChPT: THE GENERAL FRAMEWORK}

The general framework for a systematic study of radiative corrections 
in ChPT has 
been described in~ \cite{Urech,NeuRup}. It consists in writing down a 
low-momentum representation for the generating functional of QCD Green's 
functions of quark bilinears in the presence of the electromagnetic field,
\begin{equation}
e^{i{\cal Z}[v_\mu,a_\mu,s,p,Q_L,Q_R]}\,=\,
\int{\cal D}[\mu]_{QCD}{\cal D}[A_\mu]e^{i\int d^4x{\cal L}}\,,
\end{equation}
with 
\begin{equation}
{\cal L}\,=\,{\cal L}^0_{QCD}+{\cal L}^0_{\gamma}+
{\bar q}\gamma^\mu[v_\mu+\gamma_5a_\mu]q-{\bar q}[s-i\gamma_5 p]q 
+A_\mu[{\overline{q_L}}\gamma^\mu Q_Lq_L+{\overline{q_R}}\gamma^\mu Q_Rq_R].
\end{equation}
Here ${\cal L}^0_{QCD}$ is the QCD lagrangian with $N_f$ flavours of massless 
quarks, while ${\cal L}^0_{\gamma}$ is the Maxwell lagrangian of the photon 
field. The coupling of the latter to the left-handed and right-handed quark 
fields, $q_{L,R}=\frac{1\mp\gamma_5}{2}q$, occurs via the spurion sources 
$Q_{L,R}(x)$. Under local $SU(N_F)_L\times SU(N_f)_R$ chiral transformations 
$(g_L(x),g_R(x))$, they transform as (the transformation 
properties of the vector ($v_\mu$), axial ($a_\mu$), scalar ($s$) and  
pseudoscalar ($p$) sources can be found in Ref. \cite{GL2})
\begin{equation}\label{transfo}
q_{I}(x)\to g_I(x)q_I(x)\,,\ Q_{I}(x)\to g_I(x)Q_I(x)g_I(x)^+\,,\ I=L,R,
\end{equation}
so that the generating functional ${\cal Z}$ remains invariant (up to the 
usual Wess-Zumino term). Thus, although the electromagnetic interaction 
represents an explicit breaking of chiral symmetry, this breaking occurs in 
a well defined way, which is precisely the information encoded in the 
transformation properties of Eq. \ref{transfo}, much in the same way as the 
transformation properties of the scalar source $s(x)$ conveys the information 
on how the quark masses break chiral symmetry. At the end of the calculation, 
the sources $v_\mu(x)$, $a_{\mu}(x)$ and $p(x)$ are set to zero, $s(x)$ 
becomes the diagonal quark mass matrix, while the electromagnetic spurions 
are turned into the diagonal charge matrix of the quarks.
Additional symmetries of 
${\cal Z}$ consist of the discrete transformations like parity 
and charge conjugation. 
Finally, ${\cal L}$ is invariant under an additional charge conjugation type 
symmetry, which 
however affects only the photon field and the electromagnatic spurion sources,
\begin{equation}\label{discrete2}
Q_{L,R} \to -Q_{L,R}(x)\,,\ A_{\mu}(x) \to -A_{\mu}(x)\,.
\end{equation}
The low-energy representation of ${\cal Z}$ is constructed systematically in 
an expansion in powers of momenta, of quark masses and of the electromagnetic 
coupling, by computing tree and loop graphs with an effective lagrangian 
${\cal L}_{\mbox{\rm\small eff}}$
involving the $N_f\times N_f$ matrix $U(x)$ of pseudoscalar fields, and 
constrained by the chiral symmetry 
properties as well as the above discrete symmetries.

At lowest order, in the counting scheme where the electric charge $e$ and the 
spurions $Q_{L,R}(x)$ count as ${\cal O}({\mbox{\rm p}})$, the effective 
lagrangian is thus simply given by (for the notation, 
we follow~ \cite{Urech,KneUrech})
\begin{equation}
{\cal L}_{\mbox{\rm\small eff}}^{(2)}\,=\,{{F^2}\over 4}\,
\langle\,d^{\mu}U^+d_{\mu}U\,+\,
\chi^+U+U^+\chi\,\rangle
\,-{1\over 4}F^{\mu\nu}F_{\mu\nu}\,
+\,C\,\langle\,Q_{R}UQ_{L}U^+\,\rangle\,.
\end{equation}
The effect of the electromagnetic interaction is contained in the 
covariant derivative $d_\mu$, defined as 
$d_{\mu}U\ =\ \partial_{\mu}U-i(v_{\mu}+Q_{R}A_{\mu}+a_{\mu})U
+iU(v_{\mu}+Q_{L}A_{\mu}-a_{\mu})$,
and in the low-energy constant $C$, which at this order is responsible for 
the mass difference of the charged and neutral pions,
\begin{equation}
\Delta_\pi\equiv M_{\pi^{\pm}}^2-M_{\pi^0}^2\,=\,2C{e^2}/{F^2}.
\end{equation}
In fact, for the case of two light flavours ($N_f=2$), to which we restrict 
ourselves  from now on, this is the only direct effect induced by this 
counterterm. Of 
course, this mass splitting will in turn modify the kinematics of the 
low-energy $\pi\pi$ amplitudes and the corresponding scattering lengths. 
The details of this lowest order analysis can be found in Ref.~ 
\cite{KneUrech}. Here, we shall rather consider the structure of the 
effective theory at next-to-leading order. Besides the counterterms 
described by the well known low-energy constants $l_i$~ \cite{GL1}, there are 
now, if we restrict ourselves to constant spurion sources, 11 additional 
counterterms at order ${\cal O}(e^2{\mbox{\rm p}}^2)$, and three more 
at order 
${\cal O}(e^4)$. The latter contribute only to the scattering amplitudes 
involving charged pions alone. The complete list of these counterterms 
$k_i$, $i=1,...14$ and of their $\beta$-function coefficients can be found 
in Refs.~ \cite{KneUrech,Meissner}.

\section*{RADIATIVE CORRECTIONS TO THE ONE LOOP $\pi^+\pi^-\to\pi^0\pi^0$ 
AMPLITUDE}

The computation of the amplitude ${\cal A}^{+-;00}(s,t,u)$ for the process  
$\pi^+\pi^-\to\pi^0\pi^0$, including corrections of order 
${\cal O}(e^2{\mbox{\rm p}}^2)$ and of order ${\cal O}(e^4)$, is then a 
straightforward exercice in quantum 
field theory. The explicit expressions can be found in Ref.~ \cite{KneUrech} 
and will not be reproduced here. Let us rather discuss some features of the 
one-photon exchange graph of Fig. 1, which induces an electromagnetic 
correction to the strong vertex. This graph contains both the long range 
Coulomb 
interaction between the charged pions, and an infrared singularity. 
The latter is treated in the usual way, the physical, infrared finite, 
observable being the cross section for $\pi^+\pi^-\to\pi^0\pi^0$ with the 
emission of soft photons (one soft photon is enough at the order at which 
we are working here). The Coulomb force leads to a singular behaviour of 
the amplitude ${\cal A}^{+-;00}(s,t,u)$ at threshold ($q$ denotes the 
momentum of the charged pions in the center of mass frame),
\begin{equation}
Re {\cal A}^{+-;00}(s,t,u) \ =
\ -{{4M_{\pi^\pm}^2-M_{\pi^0}^2}\over{F_\pi^2}}\cdot{{e^2}\over16}\cdot
{{M_{\pi^{\pm}}}\over{q}}+Re {\cal A}^{+-;00}_{\mbox{\rm\small thr}}+O(q)\,,
\end{equation} 
with
\begin{eqnarray}
Re{\cal A}^{+-;00}_{\mbox{\rm\small thr}}
&=& 32\pi\,\left[\,-{1\over 3}(a_0^0)_{\mbox{\rm\small str}}\,
+\,{1\over 3}(a_0^2)_{\mbox{\rm\small str}}\,\right]
\,-\,{{\Delta_{\pi}}\over{F_\pi^2}}\,
+\,{{e^2M_{\pi^0}^2}\over{32\pi^2F_\pi^2}}\,(\,30-3{\cal K}^{\pm 0}_1+
{\cal K}^{\pm 0}_2\,)
\nonumber\\
&&-{{\Delta_{\pi}}\over{48\pi^2F_\pi^4}}\,
\big[\,M_{\pi^0}^2(1+4{\bar l}_1+3{\bar l}_3-12{\bar l}_4)-
6F_\pi^2e^2(10-{\cal K}_1^{\pm 0})\,\big]
\nonumber\\
&&+{{\Delta_{\pi}^2}\over{480\pi^2F_\pi^4}}\,
\big[\,212-40{\bar l}_1-15{\bar l}_3+180{\bar l}_4\,\big]\,,
\end{eqnarray}
and $(a_0^0)_{\mbox{\rm\small str}}$ and $(a_0^2)_{\mbox{\rm\small str}}$ 
denote the S wave scattering lengths in the presence of the strong 
interactions only, but expressed, for convention reasons, in terms of the 
charged pion mass~ \cite{GL1},
\begin{eqnarray}
(a_0^0)_{\mbox{\rm\small str}} &=&{{7M_{\pi^\pm}^2}\over{32\pi F_\pi^2}}\,\bigg\{\,
1+{5\over{84\pi^2}}\,{{M_{\pi^\pm}^2}\over{F_\pi^2}}\,
\bigg[\,{\bar l}_1+2{\bar l}_2-{3\over 8}{\bar l}_3+{{21}\over{10}}{\bar l}_4
+{21\over 8}\,\bigg]\,\bigg\}
\nonumber\\
(a_0^2)_{\mbox{\rm\small str}} &=&-\,{{M_{\pi^\pm}^2}\over{16\pi F_\pi^2}}\,\bigg\{\,
1-{1\over{12\pi^2}}\,{{M_{\pi^\pm}^2}\over{F_\pi^2}}\,
\bigg[\,{\bar l}_1+2{\bar l}_2-{3\over 8}{\bar l}_3-{{3}\over{2}}{\bar l}_4
+{3\over 8}\,\bigg]\,\bigg\} , 
\end{eqnarray}
corresponding to the numerical values (we use 
$F_\pi= 92.4$ MeV) $(a_0^0)_{\mbox{\rm\small str}}=0.20\pm 0.01$ and
$(a_0^2)_{\mbox{\rm\small str}}= -0.043\pm0.004$, respectively~ \cite{GL1}.
The quantity $Re{\cal A}^{+-;00}_{\mbox{\rm\small thr}}$, which is by itself 
free of the infrared divergence mentioned above, appears directly in the 
lifetime of the pionium atom~ \cite{Gall}, the long range Coulomb interaction 
being, in that case, absorbed by the bound state dynamics. The contributions 
of the low-energy constants $k_i$ are contained in the two quantities 
${\cal K}_1^{\pm 0}$ and 
${\cal K}_2^{\pm 0}$. Naive dimensional estimates lead to 
$({e^2F_{\pi}^2}/{M_{\pi^0}^2})\,{\cal K}^{\pm 0}_1\ =\ 1.8\pm 0.9$ and
$({e^2F_\pi^2}/{M_{\pi^0}^2})\,{\cal K}^{\pm 0}_2\ =\ 0.5\pm 2.2
$.
With these estimates, one obtains~  \cite{KneUrech}
\begin{equation}
{1\over{32\pi}}\,Re A^{+-;00}_{\mbox{\rm\small thr}}-
\left[\,-{1\over 3}(a_0^0)_{\mbox{\rm\small str}}\,
+\,{1\over 3}(a_0^2)_{\mbox{\rm\small str}}\,\right]
\ =\ (-1.2\pm 0.7)\times 10^{-3}\,,
\end{equation}
whereas the two-loop correction to the same combination of scattering lengths 
appearing between brackets amounts to $\sim -4\times 10^{-3}$. For a  more 
careful evaluation of the contribution of the counterterms $k_i$ to 
$Re A^{+-;00}_{\mbox{\rm\small thr}}$, see~ \cite{GasLyuRus}.

\parbox{6cm}{
\begin{center}
\epsfig{figure=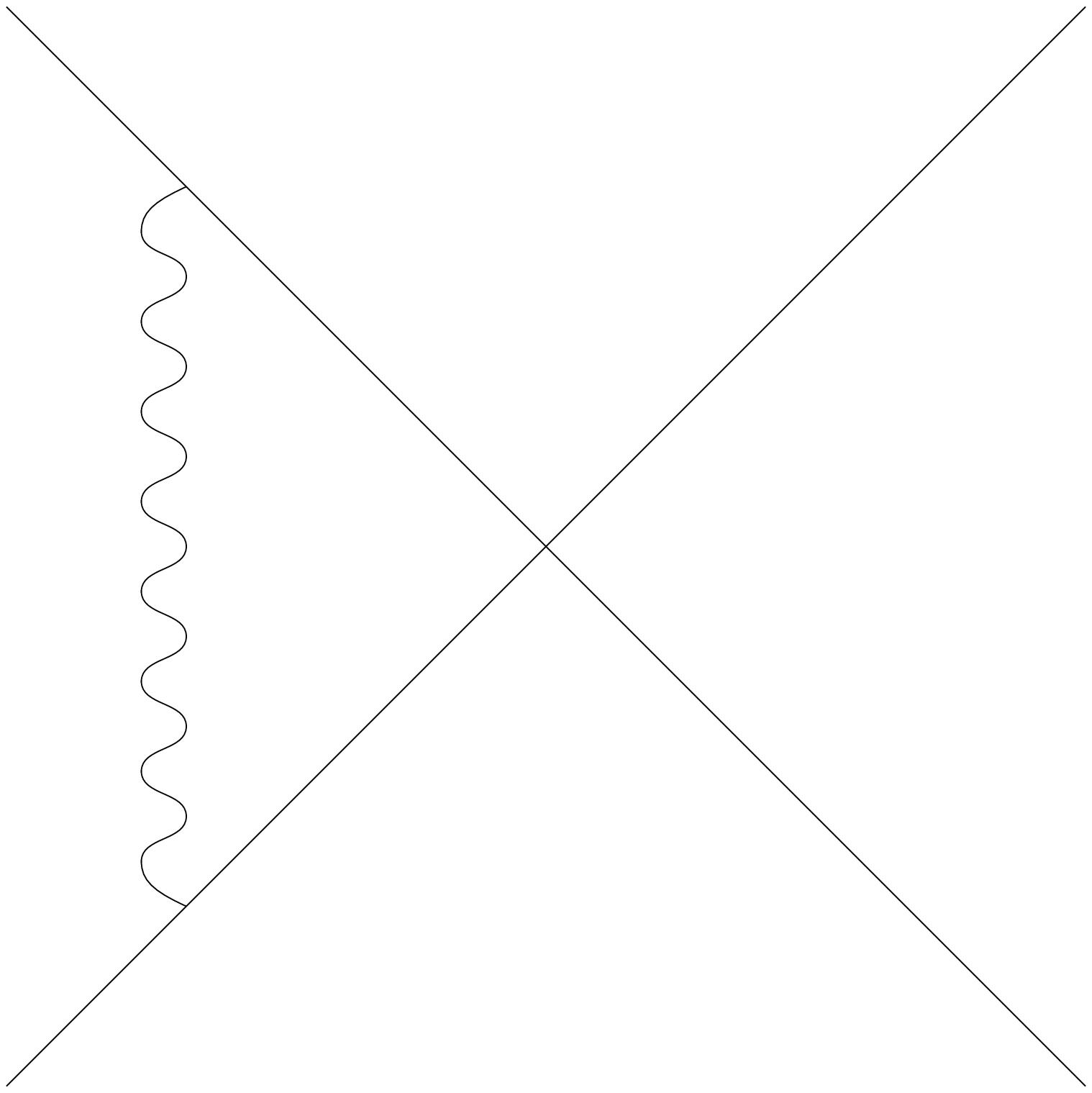,width=4.5cm,height=4.5cm,angle=-90}
\end{center}}
\parbox{5cm}{\vspace*{4.0cm}
\noindent
\parbox{5.2cm}
{\small \setlength{\baselineskip}{2.6ex} Fig.~1. The one photon exchange 
electromagnetic correction to the strong vertex.}}

\section*{CONCLUSION}

We have evaluated the radiative corrections of order 
${\cal O}(e^2{\mbox{\rm p}}^2)$ and of order ${\cal O}(e^4)$ to the amplitude 
${\cal A}^{+-;00}(s,t,u)$, 
which is relevant for the description of the pionium lifetime. Similar 
results for the scattering process involving only neutral pions can be found 
in Refs.~ \cite{KneUrech,Meissner}. The formalism 
presented here for  the case $N_f=2$ has also been applied to the study of 
radiative corrections to the pion form factors~ 
\cite{Kubis}. Radiative corrections for the scattering amplitudes involving 
only charged pions, which would be relevant for the $2p-2s$ level-shift of 
pionium, for instance, have however not been worked out so far.

Information on the low-energy scattering of pions can only be obtained in an 
indirect 
way, either from the pionium lifetime, or from $K_{\ell 4}$ decays. This last 
process, however, has electromagnetic corrections of its own, which are only 
partly covered by the present analysis. A systematic framework devoted to the 
study of radiative correction for the semi-leptonic processes has been 
presented in Refs.~ \cite{leptons,Neufeld}. 
\bibliographystyle{unsrt}

\end{document}